\documentclass[%
 reprint,
 amsmath,amssymb,
 aps,
 prb,
]{revtex4-2}

\usepackage[dvisvgm, usenames, dvipsnames]{xcolor}
\usepackage{inputenc}
\usepackage{graphicx}
\usepackage{dcolumn}
\usepackage{bm}

\begin{document}
    \preprint{APS/123-QED}
    \title{Polyvalent Machine-Learned Potential for Cobalt: from Bulk to Nanoparticles}
    \author{Marthe Bideault}
    \affiliation{Materials Design SARL, 42 avenue Verdier, 92120 Montrouge, France}
    \affiliation{%
    ICMMO/SP2M, Université Paris-Saclay, UMR 8182, 17 avenue des Sciences, 91400 Orsay, France
    }
    \author{Jérôme Creuze}
    \affiliation{%
    ICMMO/SP2M, Université Paris-Saclay, UMR 8182, 17 avenue des Sciences, 91400 Orsay, France
    }%
    \author{Ryoji Asahi}
    \affiliation{%
    Institute of Materials Innovation, Nagoya University, Nagoya 464-8603, Japan
    }%
    \author{Erich Wimmer}
    \affiliation{%
    Materials Design SARL, 42 avenue Verdier, 92120 Montrouge, France
    }%
    \date{\today}

    \begin{abstract}
	   We present the development and applications of a quadratic Spectral Neighbor Analysis Potential (q-SNAP) for ferromagnetic cobalt. Trained on Density Functional Theory calculations using the Perdew-Burke-Ernzerhof (DFT-PBE) functional, this machine-learned potential enables simulations of large systems over extended time scales across a wide range of temperatures and pressures at near DFT accuracy. It is validated by closely reproducing the phonon dispersions of hexagonal close-packed (hcp) and face-centered cubic (fcc) Co, surface energies, and the relative stability of nanoparticles of various shapes. An important feature of this novel potential is its numerical stability in long molecular dynamics simulations. This robustness is exploited to compute the heat capacity of nanoparticles containing up to 9201 atoms, showing convergence to less than 2 J.K\textsuperscript{-1}.mol\textsuperscript{-1} after 100 ns. Computations of the melting temperature of nanoparticles as a function of their size revealed a convergence to the bulk limit in excellent agreement with the experimental value. Thus, the new, highly accurate machine-learned potential for Co opens exciting opportunities for further applications such as the dynamics of nanoparticles in catalytic reactions.
    \end{abstract}

    \maketitle

    \section{\label{sec1:intro}Introduction}
    
    Elemental cobalt is a ferromagnetic transition metal that exists in two allotropic forms: at ambient conditions, the most stable phase is $\alpha$-Co, which crystallizes in a hexagonal close-packed structure. At higher temperatures, face-centered cubic $\beta$-Co becomes the most stable form. Matter \emph{et. al.} give a transition temperature of 693~K \cite{matter-tdp}. Cobalt metal is a relatively hard material with a melting point of 1768$\pm$1 K \cite{melting-p-co}. 
    
    Cobalt is used in a wide range of applications including wear-resistant high-strength alloys \cite{coutsouradis_cobalt-based_1987}, Li-ion batteries \cite{nitta2015}, magnetic recording \cite{ariake2005}, permanent magnets \cite{co-magnet}, medical implants \cite{patel2012}, dentistry \cite {kassapidou_cobalt_dental_2023}, and  radiotherapy \cite {schreiner2009}. In catalysis, Co is typically employed in the form of nanoparticles \cite{ryoji-CoBa}. 
    
    In fact, nanoparticles have revolutionized the field of catalysis with their unique size-dependent properties and high surface-to-volume ratio \cite{np-in-cata}. These nanoscale materials offer unprecedented opportunities for accelerating reactions, enhancing selectivity, and enabling greener and more efficient processes, while minimizing the amount of catalytic material. 
    
    In support of further development and optimization of catalysts, accurate atomic-scale simulations are highly valuable. Density Functional Theory (DFT) \cite{kohn-sham} is a preferred approach for systems featuring complex chemistry in a great variety of structural environments. However, these calculations are computationally demanding, limiting time scale and model size to few hundreds of picoseconds and atoms, respectively. Often, temperature effects are omitted by focusing on structures with minimum total energy. While a wealth of valuable information has been gained from this type of static ground state approaches, more realistic models are needed to include dynamic effects, especially in catalytic processes involving nanoparticles. 
    
    Interatomic potentials such as those obtained with the embedded atom method (EAM) \cite{eam} enable simulations of systems containing hundreds of thousands of atoms over time frames up to microseconds, but they are not suited for complex chemistries and substantial variations in the atomic environment. The rise of machine-learned potentials (MLPs) has opened exciting novel avenues \cite{mlp-csanyi, behler-parrinello, jmr-focus-issue}. Training these MLPs on data from \emph{ab initio} calculations yields force fields usable at the same scales as empirical potentials, while maintaining the generality and accuracy of DFT.
    
    Machine-learned potentials are founded on a descriptor of atomic environments and a model that links them to local energies. Descriptors are mathematical entities invariant under translation, rotation, and permutation of atoms of the same type. A diverse array of descriptors exists, such as the bispectrum (BSO(4)) \cite{bso4}, the Smooth Overlap of Atomic Positions (SOAP) \cite{soap}, and the Atomic Cluster Expansion (ACE) \cite{ace}. Likewise, models span from simple linear regression to multi-layer neural networks \cite{bso4, snap, efficient-cosmin, ace, nnp}.  Thompson \emph{et al.} \cite{snap} demonstrated a linear relationship between bispectrum descriptors and system observables (energy, forces and stress), leading to the introduction of the Spectral Neighbor Analysis Potential (SNAP). Later, they showed that adding quadratic terms to the model (leading to a q-SNAP) offers improved accuracy \cite{qsnap}.
    
    In this article, we present the construction of a q-SNAP for cobalt, dedicated to bulk, surface, and nanoparticle modeling. To demonstrate its quality, we show that it reproduces phonon dispersions and surface energies of hexagonal close-packed (hcp) and face-centered cubic (fcc) Co, as well as the relative stability of nanoparticles of various shapes at the accuracy of DFT. Dynamical properties including the $\alpha$-$\beta$ phase transition, thermal expansion, and the melting point are predicted in remarkable agreement with experimental values. Additionally, we compare the capabilities of this q-SNAP with those of a recent EAM potential for Co \cite{mishin} in predicting quantities that can be critical in catalysis, such as surface energies and the vacancy formation energies on nanoparticle vertices.
    
    The paper is organized as follows: in Section \ref{sec1:pot-creation}, we outline the approach used for fitting and testing the potential. In Section \ref{sec1:nps}, we validate and apply the new q-SNAP  by computing properties of bulk Co, surfaces, and nanoparticles. This includes an analysis of the relative stability of nanoparticles of different shapes as a function of their size and the determination of vacancy formation on their vertices. The suitability and robustness of the new potential to model dynamic processes is demonstrated by computing the heat capacity and the dependence of the melting temperature of nanoparticles as a function of their size. The conclusions and perspectives are presented in Section \ref{sec1:ccl}.
    
    \section{\label{sec1:pot-creation}Construction of the potential}
        
    \subsection{\label{sec2:fitting}Fitting process}
    
    \subsubsection{\label{sec3:training-set}Training set}
    
    \begin{table}
    \caption{\label{tab:db-fit} Details of the different classes used in the training set. The first column describes the class, the second one gives the number of atoms in the cells used for DFT-PBE calculations, and the three last columns correspond to the total number of energies $n_E$, forces $n_F$, and stresses $n_S$ used for the training/validation procedure. "Defects" stands for bulk supercells with interstitials and vacancies.}
    \begin{ruledtabular}
    \begin{tabular}{lcccc}
    Content & Atoms & $n_E$ & $n_F$ & $n_S$\\
    \hline
    Unit cells & 2-6 & 204/22 & 2640/252 & 1224/132\\
    MD bulk & 32-60 & 227/25 & 34110/3762 & 1362/150\\
    Defects & 55-127 & 22/2 & 6300/450 & 132/12\\
    Surfaces & 10-288 & 99/11 & 37866/3846 & 594/66\\
    Nanoparticles & 2-308 & 394/43 & 127278/13134 & 2364/258\\
    \end{tabular}
    \end{ruledtabular}
    \end{table}
    
    \begin{figure}
	   \centering
    	\includegraphics[width=4cm]{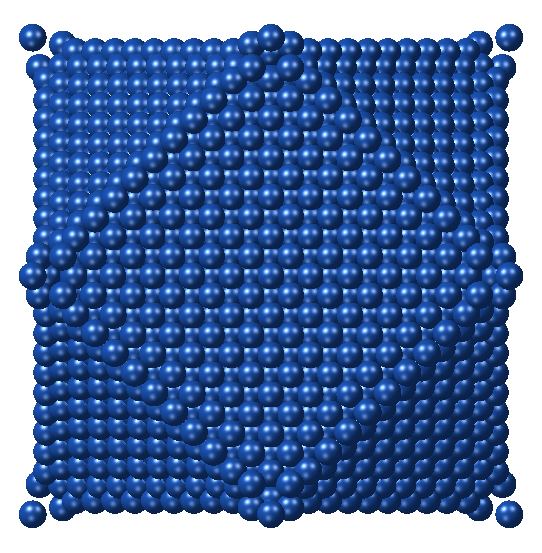}
	   \includegraphics[width=4cm]{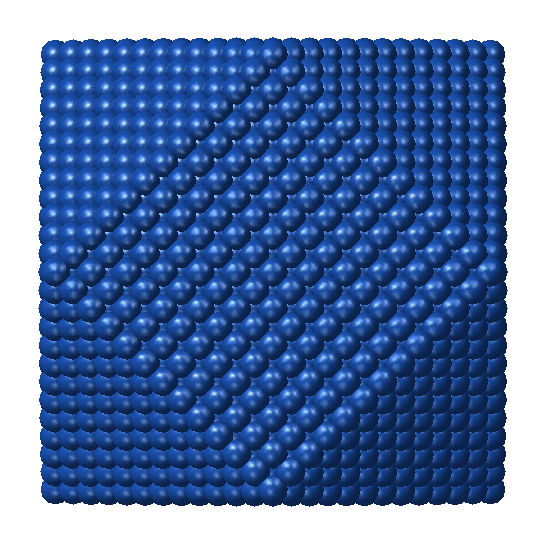}
    	\caption{Final structure from the minimization of a cobalt cuboctahedral nanoparticle of 6525 atoms initially constructed with the lattice parameter of gold (=1.15$\times$ that of cobalt), performed with a q-SNAP trained with (right) and without (left) small nanoclusters, illustrating the need to include small nanoclusters in the training set.}
	   \label{fig:oreilles}
    \end{figure}
    
    While there exists a range of generic approaches for the choice of training sets such as Bayesian methods \cite{bayesian1, bayesian2} or entropy-based active learning \cite{entropy_based_active_learning}, ultimately the choice of the training set depends on the target systems and physical phenomena such as melting, which determine the appropriate choice. Given our goal of simulating dynamic effects of bulk systems, surfaces, and especially nanoparticles, we have taken the following approach. The training set is initially created by performing \emph{ab initio} calculations on a set of structures that represent cobalt in a wide range of atomic environments. Four crystallographic phases are considered: hcp, fcc, bcc and $\omega$. Even if they are not stable for cobalt, we decided to include bcc and $\omega$ phases so that the potential learns their relative stability. Unit cells of these phases are deformed in all directions up to 1\% strain. Supercells and surfaces are also constructed. Various defects, such as interstitials and vacancies, are introduced into selected supercells, while adatoms, vacancies, and steps are incorporated into certain surface structures. Additionally, we generated icosahedra, truncated octahedra, cuboctahedra, decahedra, hcp, and spherical nanoparticles, each containing up to a few hundred atoms, with defects such as vacancies and adatoms also included. Furthermore, liquid structures are also included. These amorphous cells contain 60 atoms and are constructed at various densities ranking from 6.5 to 8.9 g.cm\textsuperscript{-3} using the amorphous builder of \textit {MedeA} \cite{medea}. Subsequently, all these structures are subjected to molecular dynamics simulations at 400, 800 and 1200 K, using the EAM potential developed by Pun and Mishin \cite{mishin}. These simulations last for 100 ps using a timestep of 2 fs. Two snapshots are included for the simulations at 400 K, whereas 3 are taken from the trajectories at 800 and 1200 K. We chose to use the EAM instead of \emph{ab initio} molecular dynamics because it allows us to create long trajectories, and thus highly uncorrelated structures.
    
    Additionally, we noticed that including small nanoclusters ranging from 2 to 30 atoms improves the potential’s robustness, as already reported in \cite{small_clusters}. Indeed, these clusters possess bispectrum descriptors that significantly differ from those of any bulk phase. In surface studies, this type of information is crucial since molecular dynamics simulations at high temperatures can easily lead to the presence of highly undercoordinated atoms, such as adatoms. A training set without these nanoclusters results in a q-SNAP that is unable to model such systems, due to an extreme increase in energy for some atoms of these special configurations (cf. figure \ref{fig:oreilles}).
    
    Finally, we included a larger number of icosahedra compared to truncated octahedra and hcp nanoparticles, given their non-crystalline structure without any equivalent in the bulk. Otherwise, they would have been underrepresented in the training set, whereas it is often the most stable morphology for transition metal nanoparticles of a few dozen to a few hundreds atoms \cite{mpnp-baletto}. We chose not to increase the representation of decahedra because they are inherently unstable for cobalt \cite{farkas} and fall outside the scope of our study. 
    
    The bispectrum descriptors are computed for all these configurations and are labeled by energy, forces, and stress obtained from \emph{ab initio} calculations. Spin-polarized single point energy calculations are performed using the Vienna Ab Initio Simulation Package (VASP) \cite{vasp1, vasp2} with the Perdew-Burke-Ernzerhof (PBE) exchange-correlation functional \cite{pbe} based on a generalized gradient approximation (GGA) \cite{gga}, as integrated in the \emph{MedeA} materials modeling environment \cite{medea}. The core electrons up to the 3\emph{p} level were frozen and their interactions with the remaining nine valence electrons were described using the projector augmented wave method (PAW) \cite{paw}. A plane-wave energy cutoff of 300 eV was used and the k-spacing in the Brillouin zone was set to 0.2 \AA\textsuperscript{-1} in the periodic directions, while only one k-point was used for non-periodic directions. The Methfessel-Paxton scheme \cite{mp-scheme} with a smearing width of 0.2 eV was employed for the integration over the Brillouin zone. To reduce the noise in the forces, an additional support grid was used for the evaluation of the augmentation charges. The SCF convergence criterion was set to 10\textsuperscript{-5} eV. To avoid interactions due to the periodic boundary conditions, a vacuum space of 15 \AA\ was set in the non-periodic directions of the cell.
    The ratios of energy \emph{vs.} forces and stresses in the training set are shown in Table. \ref{tab:db-fit} for each structure type.

    \subsubsection{\label{sec3:model-opt}Optimization of fitting parameters}
    \begin{figure}
    	\centering
	    \includegraphics{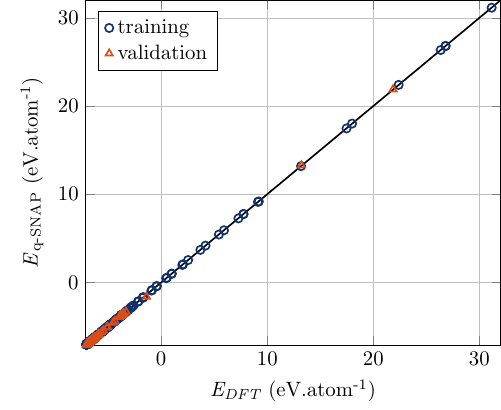}
	    \includegraphics{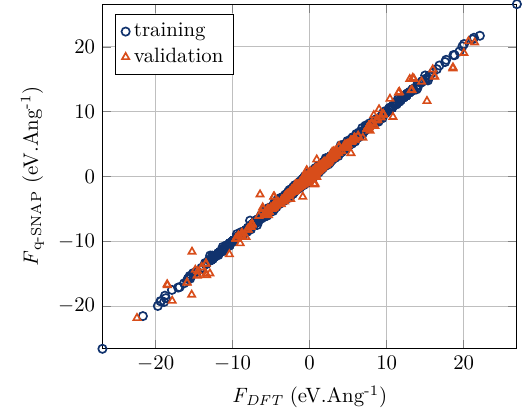}
        \caption{(Color online) q-SNAP predictions on the training observables compared with DFT-PBE for energies and forces.}
        \label{fig:fit}
    \end{figure}
    Using the Machine-Learned Potential Generator (MLPG) of the \emph{MedeA} materials modeling environment \cite{medea}, the q-SNAP model \cite{snap, qsnap} was trained on the data from DFT-PBE calculations using the least squares method. To achieve the best fit, also the band limit ($2 \times J_{max}$) and the radial cutoff needed to be optimized. To this end, a test set containing 10\% of the training configurations was extracted from the training set, and multiple q-SNAPs with different complexities were created. $J_{max}$ values of 2, 3, 4, 5 and 6 were used with radial cutoffs between 4 and 6 \AA, initially by steps of 0.2 \AA, then by steps of 0.1 \AA \ around the optimized value. For each q-SNAP, the root mean square error (RMSE) between the training and test sets was computed.
    The higher the complexity of the descriptors, the more the RMSE on the training set will be reduced. However, there comes a point where the error on the test set starts to increase again \cite{mishin-overfitting}. The optimal complexity is thus the one that minimizes the RMSE on the test set. For cobalt with our datasets, a band limit of 8 ($J_{max}=4$) and a radial cutoff of 5.0 \AA\ are the optimal values. While keeping this hyper parameters constants, weights between 0.01 and 0.1, by steps of 0.01 were tested for the forces, whereas weights of 10\textsuperscript{-5}, 10\textsuperscript{-6} and 10\textsuperscript{-7} were tested for the stresses. Finally, we used weights of 1, 0.01 and 10\textsuperscript{-6} for energies, forces and stresses, respectively.
    Fig. \ref{fig:fit} shows the training and validation errors on energy and forces of the final fit. The errors are detailed in Table \ref{tab:fit}.
    \begin{table}[b]
	   \caption{\label{tab:fit} Training and validation errors.}
	   \begin{ruledtabular}
		  \begin{tabular}{lccccc}
    		  	Type & Group & Unit & MAE & RMSE &     R\textsuperscript{2}\\
    			\hline
                Energy & train & meV.atom\textsuperscript{-1} &   2.3 & 4.3 & 1.0000\\
                Energy & test & meV.atom\textsuperscript{-1} & 8.1 & 28.6 & 0.9999\\
			    Forces & train & meV.\AA\textsuperscript{-1} & 30.9 & 54.2 & 0.9957\\
			    Forces & test & meV.\AA\textsuperscript{-1} & 33.2 & 92.7 & 0.9933\\
			    Stress & train & bar & 630 & 1434 & 1.0000\\
			    Stress & test & bar & 5931 & 41992 & 0.9998\\
		  \end{tabular}
	   \end{ruledtabular}
    \end{table}

    \subsection{\label{sec2:validation}Validation on bulk and surface properties}

    \begin{table*}
        \caption{\label{tab:ppt-hcp} Bulk properties of hcp Co.}
        \begin{ruledtabular}
		\begin{tabular}{lcccc}
			Property & Experiment & DFT-PBE
			& q-SNAP & EAM\footnotemark[1] \\ \hline 
			$a$ (\AA) & 2.507\footnotemark[2] & 2.4902 & 2.4899 & 2.5187\\
			
			$c/a$ & 1.623\footnotemark[2] & 1.6160 & 1.6149 & 1.6103\\
			
			$C_{11}$ (GPa) & 293\footnotemark[3], 319.50\footnotemark[4], 307.1\footnotemark[5] & 386 & 366 & 311.9\\
			
			$C_{12}$ (GPa) & 143\footnotemark[3], 166.09\footnotemark[4], 165.0\footnotemark[5] & 151 & 174 & 146.9\\
			
			$C_{13}$ (GPa) & 90\footnotemark[3], 102.09\footnotemark[4], 102.7\footnotemark[5] & 114 & 118 & 119.6\\
			
			$C_{33}$ (GPa) & 339\footnotemark[3], 373.60\footnotemark[4], 358.1\footnotemark[5] & 403 & 425 & 359.4\\
			
			$C_{44}$ (GPa) & 78\footnotemark[3], 82.41\footnotemark[4], 75.5\footnotemark[5] & 97 & 74 & 91.7\\
			
			$E^f_v$ (eV) & 1.4\footnotemark[6], 1.38\footnotemark[7] & & 1.8222 & 1.49 \\

            $C_V$ (298.15 K) (J.K\textsuperscript{-1}.mol\textsuperscript{-1}) & 24.732\footnotemark[8] & 22.9263\footnotemark[9] & 22.9736\footnotemark[9] & 22.4766\footnotemark[9]\\ 
   
			$T_{hcp-fcc}$ (K) & 690\footnotemark[10], 695\footnotemark[11], 700\footnotemark[12] & & 757 & 717\footnotemark[1] \\
		\end{tabular}
	    \end{ruledtabular}
	    \footnotetext[1]{Ref. \onlinecite{mishin}}
	    \footnotetext[2]{Ref. \onlinecite{a-hcp}}
	    \footnotetext[3]{Ref. \onlinecite{cij-hcp-1}}
        \footnotetext[4]{Ref. \onlinecite{cij-hcp-2}}
        \footnotetext[5]{Ref. \onlinecite{cij-hcp-3}}
        \footnotetext[6]{Ref. \onlinecite{surfE-vac-hcp}}
        \footnotetext[7]{Ref. \onlinecite{vac-hcp}}
        \footnotetext[8]{Ref. \onlinecite{cv}}
        \footnotetext[9]{Computed within the harmonic approximation}
        \footnotetext[10]{Ref. \onlinecite{all_thermal_exp}}
        \footnotetext[11]{Ref. \onlinecite{tdp-695}}
        \footnotetext[12]{Ref. \onlinecite{tdp-700}}
    \end{table*}

    \begin{table*}
        \caption{\label{tab:ppt-fcc} Bulk properties of fcc Co.}
        \begin{ruledtabular}
		\begin{tabular}{lcccc}
			Property & Experiment & DFT-PBE
			& q-SNAP & EAM\footnotemark[1]\\ \hline 
			$a$ (\AA) & 3.5447\footnotemark[2], 3.568\footnotemark[3] & 3.5103 & 3.5116 & 3.5642\\
			
			$C_{11}$ (GPa) & 260\footnotemark[4], 225\footnotemark[5], 223\footnotemark[6] & 302 & 328 & 275.7\\
			
			$C_{12}$ (GPa) & 160\footnotemark[4], 160\footnotemark[5], 186\footnotemark[6] & 173 & 201 & 158.9\\
			
			$C_{44}$ (GPa) & 110\footnotemark[4], 92\footnotemark[5], 110\footnotemark[6] & 149 & 138 & 108.2\\
			
			$E^f_v$ (eV) & 1.34 - 1.91\footnotemark[7] & 2.34\footnotemark[8], 1.71\footnotemark[8] & 1.7542 & 1.56 \\

            $C_V$ (298.15 K) (J.K\textsuperscript{-1}.mol\textsuperscript{-1}) & 24.811\footnotemark[9] & 23.1599\footnotemark[10] & 23.2380\footnotemark[10] & 
            22.6667\footnotemark[9]\\ 
			
			$T_m$ (K) & 1768\footnotemark[11], 1770\footnotemark[12] & & 1695 +/- 15 & 1898\footnotemark[1]\\
        \end{tabular}  
        \end{ruledtabular}
        \footnotetext[1]{Ref. \onlinecite{mishin}}
        \footnotetext[2]{Ref. \onlinecite{a-fcc1}}
        \footnotetext[3]{Ref. \onlinecite{a-fcc2}}
        \footnotetext[4]{Ref. \onlinecite{cij-fcc-1}}
        \footnotetext[5]{Ref. \onlinecite{cij-fcc-2}}
        \footnotetext[6]{Ref. \onlinecite{cij-fcc-3}}
        \footnotetext[7]{Ref. \onlinecite{vacE-fcc}}
        \footnotetext[8]{Ref. \onlinecite{vacE-fcc-dft}}
        \footnotetext[9]{Ref. \onlinecite{cv}}
        \footnotetext[10]{Computed within the harmonic approximation}
        \footnotetext[11]{Ref. \onlinecite{kittel}}
        \footnotetext[12]{Refs. \onlinecite{tdp-695} and \onlinecite{tdp-700}}
    \end{table*}

    We assessed the capability of the q-SNAP to accurately reproduce results obtained from periodic DFT-PBE calculations. To achieve this, numerous properties were computed and analyzed. The computational methodologies employed to calculate these properties are detailed below. DFT-PBE calculations were carried out using the parameters outlined in section \ref{sec2:fitting}. For structure optimizations, we used the conjugate gradient algorithm with a convergence criterion of 0.02 eV.\AA\textsuperscript{-1}. The LAMMPS molecular dynamics package \cite{lammps} was employed to obtain the results based on the q-SNAP. Structure minimizations were also performed using the conjugate gradient algorithm, with a convergence criterion of 10\textsuperscript{-3} eV.\AA\textsuperscript{-1}. During the molecular dynamics simulations, a time step of 2 fs was employed, and the Nosé-Hoover thermostat and barostat were used with dampings of 200 and 2000 fs to control temperature and pressure, respectively.

    Tables \ref{tab:ppt-hcp} and \ref{tab:ppt-fcc} compare the experimental values of hcp and fcc cobalt with those obtained with q-SNAP, DFT-PBE and EAM \cite{mishin}.
    The q-SNAP results shown in these tables reflect the accuracy of the DFT-PBE level of theory for calculations at $T = 0$ K. The EAM potential was fitted to reproduce experimental data, typically measured at ambient temperature. This explains the better agreement of the EAM results for structural and elastic properties with experimental data compared with the pure DFT-PBE results. However, fitting to specific experimental data runs the risk of introducing uncontrolled errors when using such a potential to compute other properties such as the temperature of phase transitions or energies.   

    \begin{figure}
        \centering
        \includegraphics[width=8.6cm]{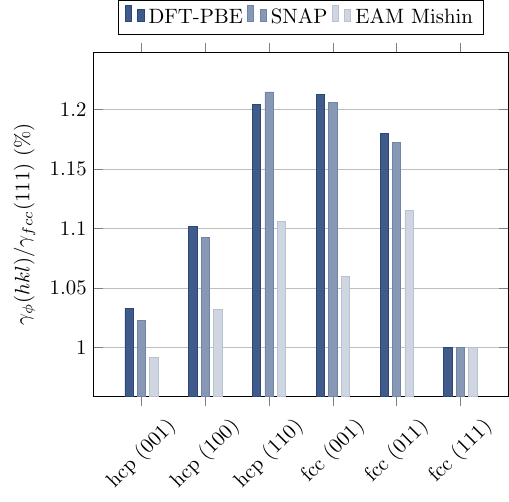}
        \caption{(Color online) Surface energies relative to the fcc (111) surface energy, calculated with the present q-SNAP and compared with those obtained by DFT-PBE and the EAM potential \cite{mishin}.}
        \label{fig:surfaces}
    \end{figure}
    The surface energy of phase $\phi$ with Miller indices $(h, k, l)$  is expressed as
    \begin{eqnarray}
        \gamma_\phi(hkl) = \frac{E_{slab}-E_{bulk_\phi} \times n_{slab}}{2A}
    \end{eqnarray}
    were $E_{slab}$ corresponds to the total energy of the slab of area of $A$ containing $n_{slab}$ atoms, and $E_{bulk_\phi}$ to the energy per atom of the minimized bulk cell of phase $\phi$.
    Slabs were constructed with a vacuum spacing of 15 \AA\  and a minimum thickness of 16 \AA. 
    As illustrated in Fig. \ref{fig:surfaces},  q-SNAP predicts surface energies for low index orientations that are extremely close to the DFT-PBE values, with a maximum difference of 47 mJ.m\textsuperscript{-2} for $\gamma_{hcp}(110)$, whereas the EAM potential leads to much larger deviations, up to 272 mJ.m\textsuperscript{-2} for the $\gamma_{fcc}(111)$. Moreover, the EAM potential finds that the hcp (001) surface is the most stable instead of the fcc (111), and predicts that $\gamma_{fcc}(011) < \gamma_{fcc}(001) < \gamma_{hcp}(110)$. The disparity observed between the EAM potential and the q-SNAP (and DFT-PBE) likely arises from the decisions made during the calibration of the EAM potential, primarily aimed at accurately reproducing bulk properties, as mentioned above. It should be mentioned, however, that the PBE level of theory tends to underestimate surface energies of transition metals \cite{surf-nrj-pbe}.  

    \begin{figure}
        \centering
        \includegraphics{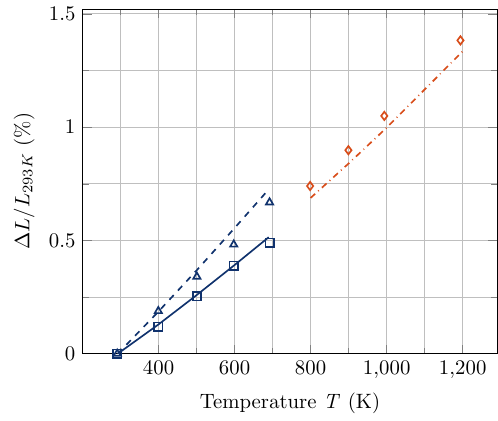}
        \caption{(Color online) Thermal expansion of hcp and fcc cobalt measured experimentally (lines) and calculated with the q-SNAP (marks). The blue solid line and squares represent the $a$ lattice parameter of hcp cobalt, whereas the dashed line and triangles refer to the $c$ one. The red dash-dotted line and diamonds represent the lattice parameter of fcc cobalt.}
        \label{fig:thermal_exp}
    \end{figure}
    A further test of the present q-SNAP is the computation of the thermal expansion of hcp and fcc phases. To do this, we ran molecular dynamics simulations in the NPT ensemble at atmospheric pressure between $T=298$ and $1200$ K, with 5x5x5 and 4x4x4 supercells for the hcp and fcc phases respectively. The temperature was incremented linearly in steps of 100 K with an equilibration of 50 ps for each temperature. Only the cell angles were fixed.
    As illustrated in Fig. \ref{fig:thermal_exp}, the q-SNAP results are in very good agreement with the experimental data from Ref. \onlinecite{all_thermal_exp}, showing that this interatomic potential reproduces well this property. The small underestimation of the thermal expansion at elevated temperatures may be due in part to the presence of vacancies in the experimental samples while the simulations are preformed for vacancy-free systems.

    \begin{figure}
        \centering
        \includegraphics[width=4.1cm]{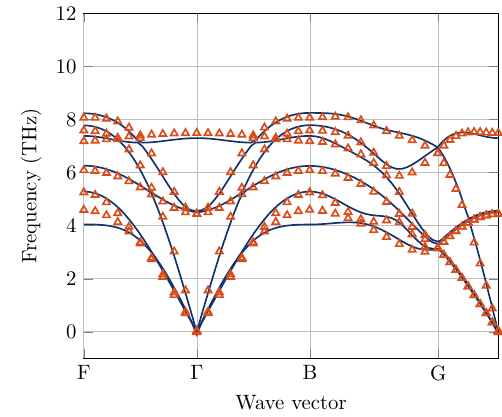}
        \includegraphics[width=4.1cm]{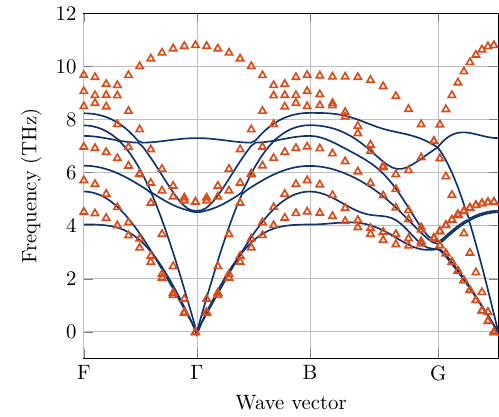}
        \includegraphics[width=4.1cm]{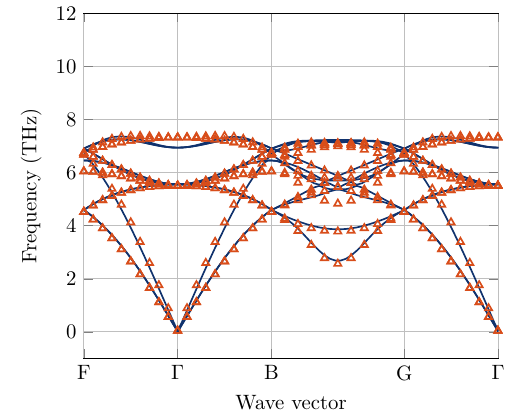}
        \includegraphics[width=4.1cm]{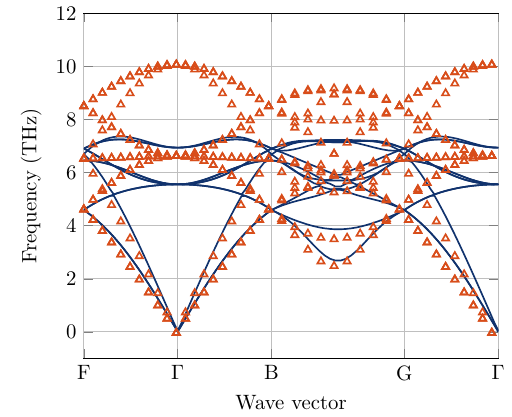}
        \caption{(Color online) Phonon dispersion curves of hcp (top) and fcc (bottom) cobalt, computed using DFT-PBE (solid lines), the q-SNAP (squares) and the EAM\cite{mishin} (triangles). Note that these results are obtained in the P1 space group.}
        \label{fig:phonon_disp}
    \end{figure}
    Phonon dispersions were computed using the phonon module as available in \emph{MedeA} \cite{medea}, which uses the method described in Ref. \onlinecite{phonon}. The dispersions predicted by the q-SNAP are in very good agreement with those predicted by DFT-PBE, as illustrated in Fig. \ref{fig:phonon_disp}, with the highest frequencies being around 8 THz for the hcp phase, consistent with experimental data \cite{phonon-exp} while the EAM of Pun \emph{et. al.} \cite{mishin} predicts the highest frequencies to be nearly 11 THz.

    \begin{figure}
        \centering
        \includegraphics{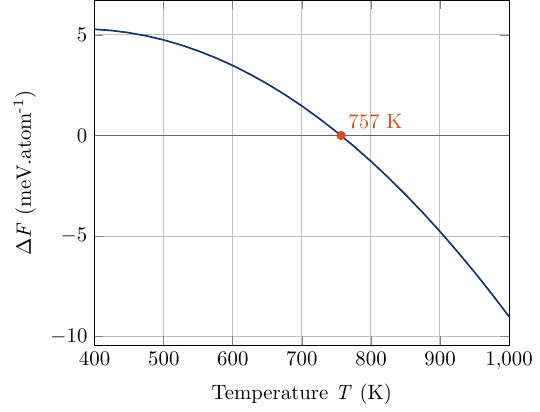}
        \caption{(Color online) Difference in Helmholtz free energy between the fcc and hcp phases of cobalt as a function of temperature. The predicted phase transition temperature at atmospheric pressure is $T = 757$ K (red dot).}
        \label{fig:tdp}
    \end{figure}
    An essential benchmark for assessing the efficiency of the potential lies in the computation of the $\alpha$-$\beta$ phase transition. Assuming that the entropic terms are solely vibrational, the Helmholtz free energy $F(T)$ of a given phase $\phi$ is defined as:
    \begin{eqnarray}\label{eq:F}
        F_\phi(T) = U(T) + E_{vib,QC}(T) + ZPE - S_{vib}(T)T
    \end{eqnarray}
    where $T$ is the temperature, $U$ the internal energy of the system, $E_{vib,QC}$ the quantum contribution of the vibrational potential energy, $ZPE$ the zero point energy and $S_{vib}$ the vibrational entropy. The computations of the three last terms rely on the vibrational density of states $DOS(\nu)$, determined via the Fourier transform of the integrated velocity auto-correlation function \cite{mdphonon1, mdphonon2}. Supercells of 1000 and 1008 atoms for the hcp and fcc phases respectively are equilibrated at a given temperature for 170 ps within the NPT ensemble. Subsequently, the velocity auto-correlation function is calculated during two simulations of 25 ps in the NVE ensemble. The Helmholtz free energy of both phases is computed using equation \ref{eq:F}, followed by polynomial fitting of the resultant curves to determine $\Delta F_{fcc-hcp}(T)$. As shown in Fig. \ref{fig:tdp}, this function intersects zero at $T = 757$ K, demonstrating rather good agreement with the experimental data, considering that a shift of only 1.6 meV.atom\textsuperscript{-1} would yield a transition temperature closer to the experimental value of 693 K \cite{matter-tdp}. Indeed, there might be uncertainties also in the experimental value.

    Last, the biphased cell method was used to estimate the melting temperature of fcc Co. Following the procedure in \cite{melting-p-method}, a 5x5x20 supercell was equilibrated for 100 ps at 800 K using NPT molecular dynamics, with only the angles fixed. Subsequently, all atoms with a fractional coordinate $z$ greater than 0.5 were fixed. A 50 ps NPT molecular dynamics simulation was performed at 3000 K to rapidly transform half of the system into a liquid state. During this simulation, only the $z$-direction of the cell was allowed to relax. Then, while still keeping only the $z$-direction free, a third NPT molecular dynamics simulation was conducted at the experimental melting point of 1768 K \cite{melting-p-co}. This simulation lasted only 1 ps, which was sufficient to equilibrate the density at the presumed melting temperature without recrystallization. Once this biphased cell was constructed, NPT molecular dynamics simulations of 200 ps were performed at various temperatures. All directions were allowed to relax but the angles were fixed. The melting temperature was determined based on the final state of the final structure. All simulations were conducted at atmospheric pressure. The q-SNAP predicts a melting temperature of 1695 +/- 15 K, which is only 73 K below the experimental value.

    In summary of this validation, the present q-SNAP exhibits excellent accuracy in reproducing a broad range of static and dynamic properties of both bulk and surfaces of hcp and fcc cobalt, thus highlighting its robustness and reliability. The agreement with experimental data such as the melting point demonstrates that the underlying level of \emph{ab initio} theory, namely DFT-PBE, provides a remarkably good description of the interatomic interactions in this system.

    \section{\label{sec1:nps}Application to nanoparticles}
    In the following analysis, our objective was to exploit the accuracy and robustness of the present q-SNAP in describing the structure and energetics of nanoparticles. This capability holds substantial importance, particularly in the context of catalysis, where accurate predictions are crucial for understanding and optimizing catalytic processes.

    Atoms involved in chemical reactions exist in states far from their equilibrium, adopting configurations of high energy. In heterogeneous catalysis, these reactions take place on surfaces or nanoparticles, where the reactants adsorb on  specific sites, which can be an adatom or any other surface pattern with reactive surface atoms \cite{boudart}. In fact, surface atoms, especially on nanoparticles, dynamically participate in catalytic reactions \cite{nh3-parrinello}.

    As a result, it is crucial to accurately characterize these surfaces and adsorption sites to predict the catalyst's surface state as a function of temperature. 

    In this part, we will show the capabilities of the present q-SNAP in modeling cobalt nanoparticles, both statically through the relative stabilities of different nanoparticle shapes and the vacancy formation on vertices, and dynamically by examining the melting process.

    \subsection{\label{sec2:stab-nps}Morphological stability at \emph{T}=0 K}
    \begin{figure}
        \centering
        \includegraphics[width=8.6cm]{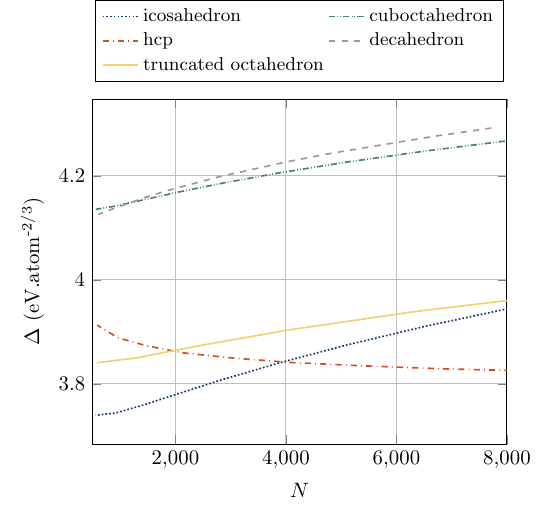}
        \caption{(Color online) Relative stability of nanoparticles of different shapes as a function of their size at 0 K. The most stable nanoparticles have the lowest $\Delta$ value.}
        \label{fig:nps_0K}
    \end{figure}
    In terms of the relative stabilities among different shapes, the present q-SNAP predicts the icosahedron to be the most stable shape up to 4000 atoms, after which the hexagonal close-packed (hcp) shape becomes the most stable. This is consistent with experimental observations of icosahedra for small nanoparticles \cite{exp-ico-np} and hcp structures for larger ones \cite{exp-hcp-np}. The truncated octahedron is more stable than the hcp up to 1900 atoms. The decahedron and cuboctahedron are considerably less stable, with a difference of 0.5 to 1 eV.atom\textsuperscript{-1} compared to the three shapes mentioned before. Their curves of energies \emph{vs.} size are nearly parallel, intersecting smoothly at 1300 atoms, where the cuboctahedron becomes more stable than the decahedron. We used the parameter $\Delta$, as introduced in Ref. \cite{mpnp-baletto}, to depict the relative stabilities in Fig. \ref{fig:nps_0K}. This choice offers a clearer distinction between the various transitions compared to merely plotting linear regressions. The parameter $\Delta$ is defined as
    \begin{equation}
        \Delta (N) = \frac{E(N) - N\epsilon_{coh}}{N^{2/3}}
    \end{equation}
    where $E(N)$ is the total energy of a nanoparticle containing $N$ atoms, and $\epsilon_{coh}$ the cohesive energy of bulk hcp cobalt, which is -7.04 eV.atom\textsuperscript{-1}.

    Farka\v{s} and de Leeuw \cite{farkas} used DFT-PBE to minimize nanoparticles of different shapes up to 1000 atoms. They determined the relative stabilities of larger systems using a linear extrapolation. Their predictions are comparable to the results of the present explicit calculations, but there are noticeable differences:  Farka\v{s} and de Leeuw \cite{farkas} predict the icosahedron to be the most stable shape up to 5500 atoms, followed by the hcp structure for larger nanoparticles. The present work, without any extrapolation, predicts the transition at 3880 atoms. In Ref. \cite{farkas}, the transition between the truncated octahedron and hcp occurs near 500 atoms while the present results give 1879 atoms. In Ref. \cite{farkas}, the curves of the decahedron and cuboctahedron are separated by minimal differences in energy, but do not intersect, while the explicit q-SNAP computations predict a transition at 1315 atoms, as can be seen from Fig. \ref{fig:nps_0K}.

    For comparison, the EAM potential \cite{mishin} predicts the icosahedron to be the most stable shape up to 2666 atoms, with the transition between the truncated octahedron and hcp occurring near 387 atoms, which is consistent with the present results. However, the stability of the cuboctahedron is closer to that of the icosahedron, truncated octahedron, and hcp than to that of the decahedron. For small nanoparticles below 400 atoms, the EAM predicts the cuboctahedron to be even more stable than hcp, in disagreement with DFT-PBE calculations and the present q-SNAP results.

    \subsection{\label{sec2:vac-vertex}Vacancy formation energy on vertices}
    \begin{figure}
        \centering
        \includegraphics[width=8.6cm]{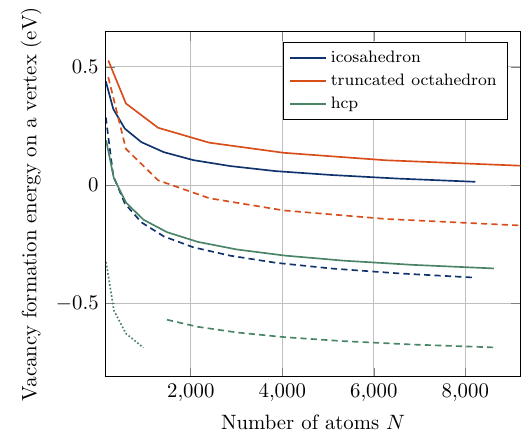}
        \caption{(Color online) Vacancy formation energy on a vertex of icosahedra, truncated octahedra and hexagonal close packed nanoparticles according to their size. Results obtained with q-SNAP and EAM are drawn in solid and dashed lines, respectively. The green dotted line corresponds to the sizes where the EAM potential predicts the edge to be shifted in order to fill the vacancy, as illustrated in Fig. \ref{fig:min-hcp}.}
        \label{fig:vac-vertex-size}
    \end{figure}
    \begin{figure}
        \centering
        \includegraphics[width=8.6cm]{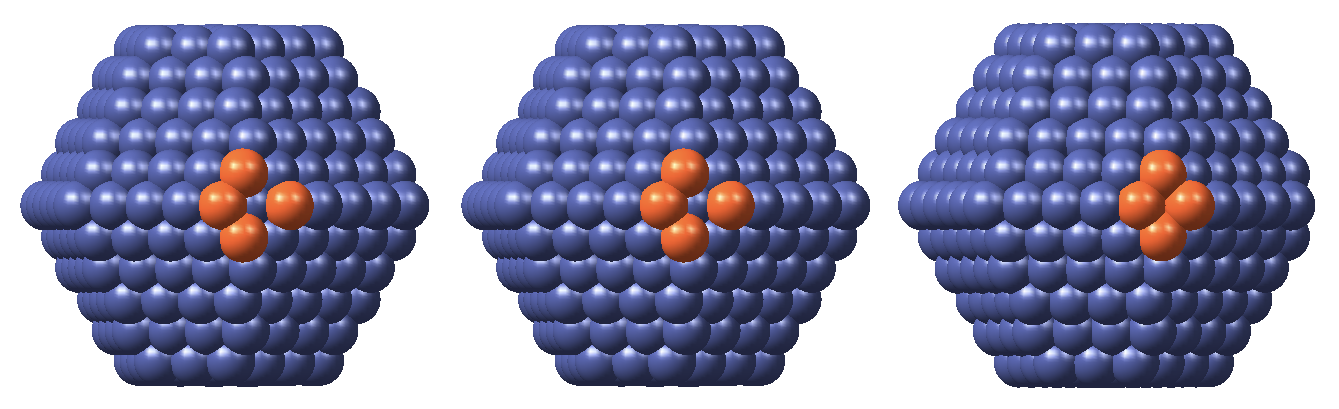}
        \caption{(Color online) Hexagonal close-packed nanoparticle (587 atoms) minimized using DFT-PBE (left), the q-SNAP (middle) and the EAM potential (right). Both q-SNAP and DFT-PBE predict minor deviations relative to the initial state (with one vertex removed from the minimized nanoparticle), wherein atoms surrounding the vacancy tend to move away from it. Conversely, the EAM potential predicts an edge shift to fill the vacancy.}
        \label{fig:min-hcp}
    \end{figure}
    \begin{figure}
        \centering
        \includegraphics[width=8.6cm]{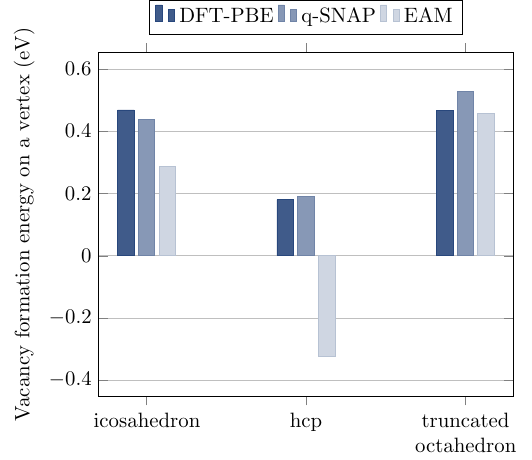}
        \caption{(Color online) Vacancy formation energy on vertices of icosahedra, hexagonal close-packed and truncated octahedra of 147, 153 and 201 atoms respectively, calculated using DFT-PBE (dark blue), the q-SNAP (blue) and the EAM potential \cite{mishin} (light blue).}
        \label{fig:verif-vertices}
    \end{figure}

    Vacancy formation energies provide a sensitive test for the quality of interatomic potentials. To this end, a comparison was made between results from q-SNAP and the EAM potential as a function of the nanoparticle's size, focusing on the vertices of nanoparticles.

    The vacancy formation energy is defined as
    \begin{eqnarray}
        \Delta E_{\text{form,vac}} = E_{\text{tot,vac}} + (N^{-1}-1)E_{\text{tot,clean}}
    \end{eqnarray}
    where $E_{\text{tot,clean}}$ is the total energy of the pristine nanoparticle without a vacancy, $E_{\text{tot,vac}}$ is the total energy of the nanoparticle with a vacancy, and $N$ is the number of atoms in the pristine nanoparticle \cite{mottet-vac}. 

    According to both potentials, the vacancy formation energy decreases with an increasing number of atoms, as shown in Fig. \ref{fig:vac-vertex-size}. The EAM potential predicts that forming a vacancy on a nanoparticle's vertex is always favorable, except for very small truncated octahedra and hcp nanoparticles, whereas the q-SNAP predicts this behavior only for hcp nanoparticles containing more than 323 atoms. For all sizes and shapes, the present q-SNAP predicts that the atoms surrounding the vacancy leave more space for it, while the EAM potential predicts the surrounding atoms to fill the vacancy. For hcp nanoparticles up to 967 atoms, the EAM potential even predicts that the entire edge is shifted to fill the vacancy, as illustrated on the right hand side of Fig. \ref{fig:min-hcp}. For larger sizes, the edge does not shift, as predicted on the whole range according to the q-SNAP and for the hcp nanoparticle of 153 atoms according to DFT-PBE. The two types of minimizations are represented by two distinct curves in Fig. \ref{fig:vac-vertex-size}. In addition, Fig. \ref{fig:verif-vertices} illustrates the excellent agreement between the q-SNAP and DFT-PBE predictions for small nanoparticles.

    \subsection{\label{sec2:nps-dynamics}Dynamics}
    
    \subsubsection{\label{sec3:stability}Stability over long simulation times}
    \begin{figure}
        \centering
        \includegraphics[width=8.6cm]{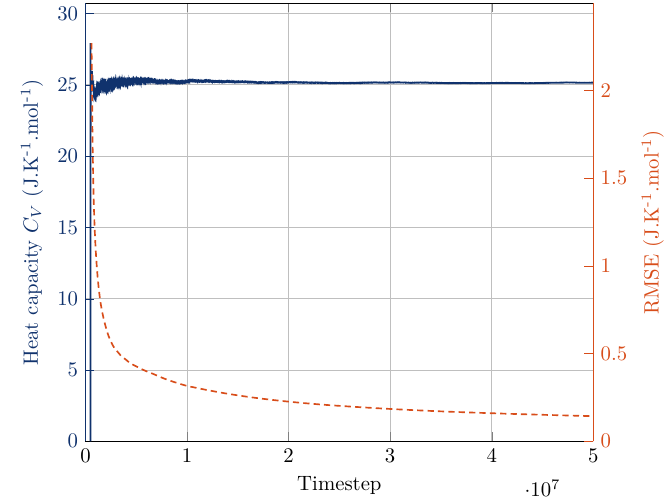}
        \caption{(Color online) Heat capacity of a 2406 atoms truncated octahedron during a 100 ns simulation at 300 K. The dashed line corresponds to the evolution of its RMSE over time.}
        \label{fig:stability_Cv}
    \end{figure}
    \begin{figure}
        \centering
        \includegraphics[width=2.7cm]{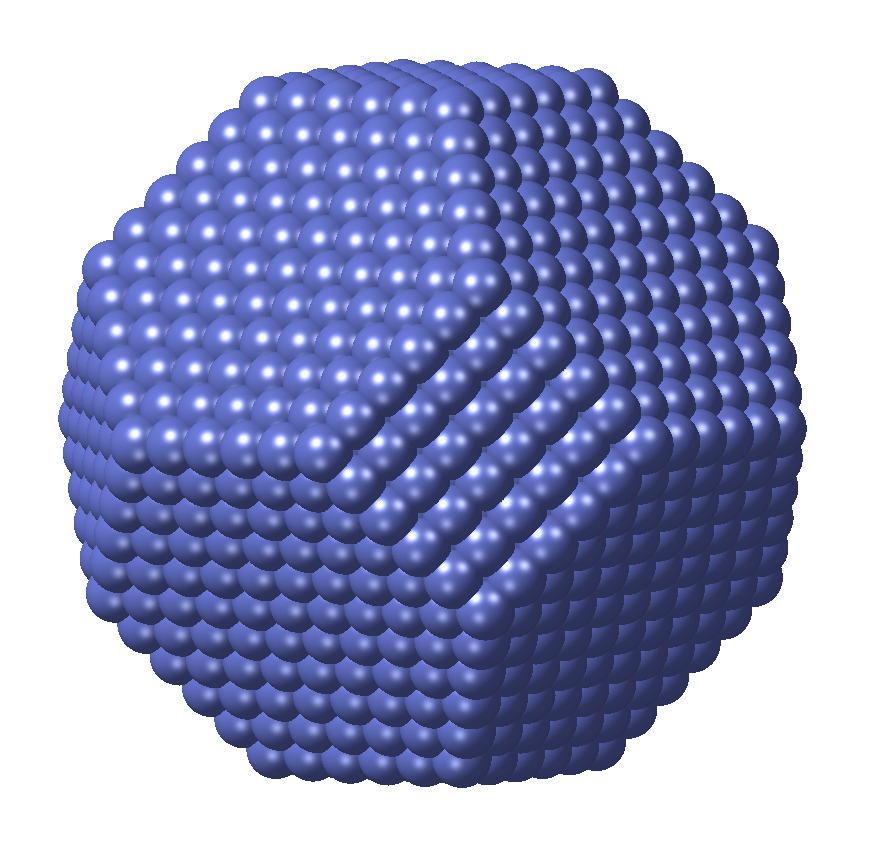}
        \includegraphics[width=2.7cm]{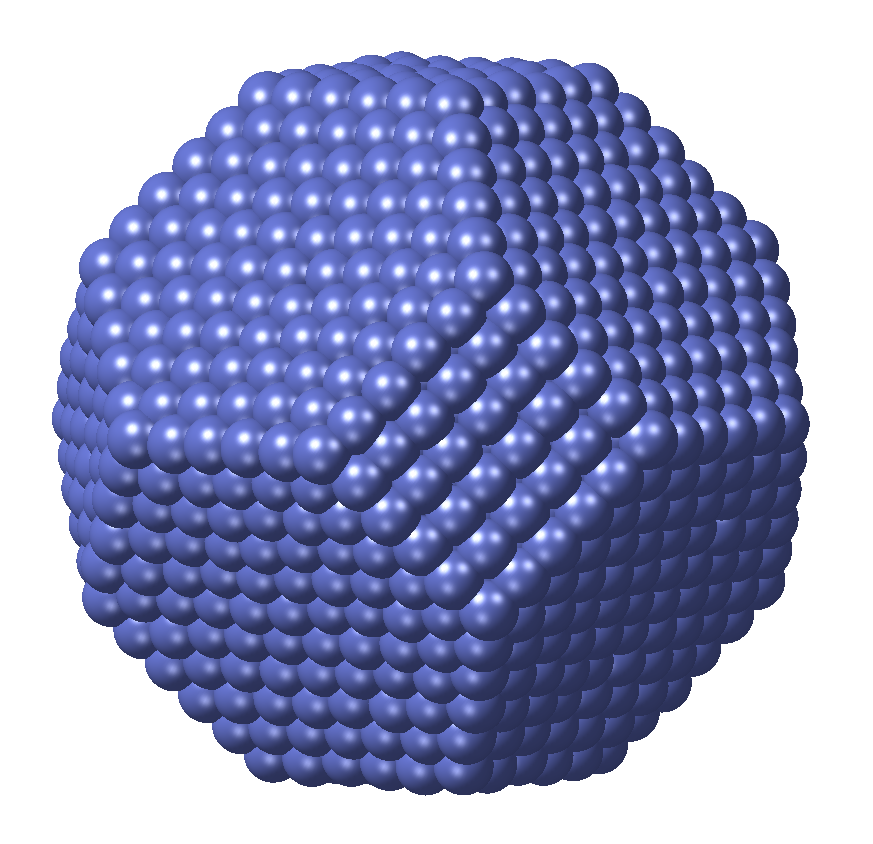}
        \includegraphics[width=2.7cm]{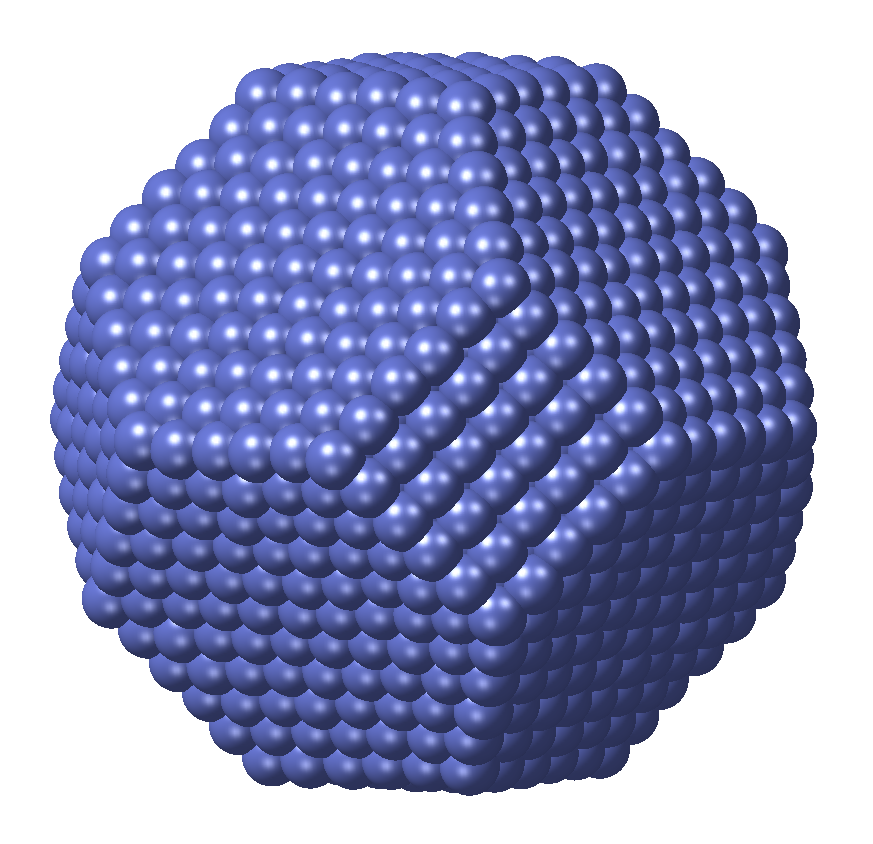}
        \caption{(Color online) Truncated octahedron of 2406 atoms at the beginning (left), after 50 ns (middle) and at the end (right) of a 100 ns simulation at 300 K.}
        \label{fig:stability_pic}
    \end{figure}
    The main purpose of this potential is to model large complex systems over timescales that are inaccessible with DFT.
    
    MLPs, particularly Neural Network Potentials (NNPs), are known to become unstable or inaccurate after a certain simulation time \cite{fail_nn}. To test the stability of the new q-SNAP, we ran a simulation of 100 ns on a 2406 atoms truncated octahedron. The nanoparticle was placed in a cubic cell with 50 \AA \ of vacuum around it and evolved through Langevin dynamics at a constant temperature of 300 K, with a timestep of 2 fs. The instantaneous total energy of the system is written every 20 fs.

   From these data, the heat capacity is computed as     
    \begin{equation}
        C_V = \frac{\langle E^2 \rangle -\langle E\rangle ^2}{N_Ank_BT^2}
    \end{equation}
    where $E$ is the total energy of the system, $N_A$ Avogadro's number, $n$ the number of atoms in the nanoparticle (here 2406), $k_B$ the Boltzmann's constant and $T$ the temperature (here 300 K). The heat capacity, resulting from the average of the values computed between 50 and 100 ns is 25.14 J.K\textsuperscript{-1}.mol\textsuperscript{-1}. This value is 9.4\% larger than that for bulk hcp Co (see Table \ref{tab:ppt-hcp}). The larger specific heat of nanoparticles is due to the lower frequency of vibrational modes of surface atoms compared to bulk atoms \cite{jerome_cv}.
    
    Fig. \ref{fig:stability_Cv} shows its constancy throughout the entire simulation, as well as its RMSE, which decreases over time, as expected. Additionally, Fig. \ref{fig:stability_pic} shows the nanoparticle at the beginning, in the middle, and at the end of the simulation, always maintaining the same shape. The remarkable stability of the presented q-SNAP is a key strength.
    
    \subsubsection{\label{sec3:melting-nps}Melting}
    \begin{figure}
        \centering
        \includegraphics[width=8.6cm]{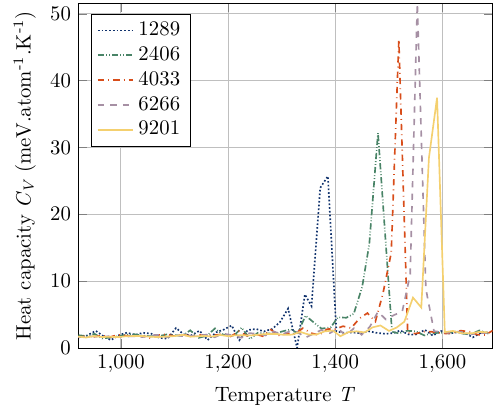}
        \caption{(Color online) Heat capacities as a function of temperature for five truncated octahedra of increasing size. The maximum of the peak corresponds to the melting temperature.}
        \label{fig:mpnp_cv}
    \end{figure}
    \begin{figure}
        \centering
        \includegraphics[width=8.6cm]{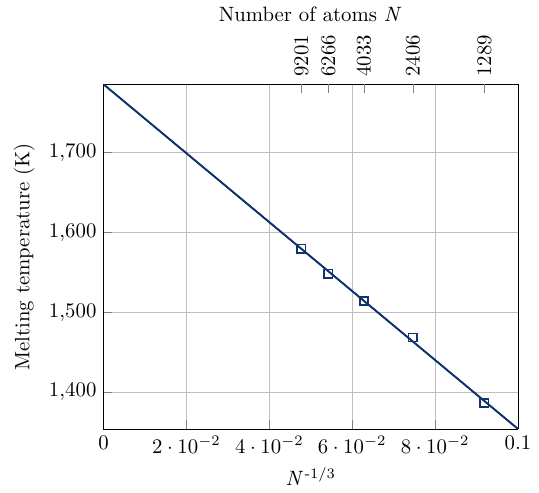}
        \caption{(Color online) Linear fit (solid line) of truncated octahedral cobalt nanoparticles melting temperatures (squares) as a function of $N^{-1/3}$, where $N$ is the number of atoms. It leads to a bulk melting point of 1785 K, with a R-squared coefficient of 0.999.}
        \label{fig:mp}
    \end{figure}

    In atomistic simulations, it is possible to calculate the melting temperature of a material using the biphased cell method, as described above. It has also been demonstrated, both experimentally \cite{mpnp-exp} and computationally \cite{mpnp-simu}, that it is possible to estimate the material's melting temperature by melting nanoparticles of varying sizes. As the size increases, the nanoparticle's melting temperature tends to that of the bulk. This phenomenon is described by the Gibbs-Thomson relation \cite{mpnp-exp}:
    \begin{equation}
        T_{m,NP} = T_{m,b} \left(1 - \frac{C}{D}\right)
    \end{equation}
    where $T_{m,NP}$ is the melting temperature of the nanoparticle with diameter $D$, $T_{m,b}$ is that of the bulk, and $C$ is a constant depending on the considered material. Thus, assuming the nanoparticle to be spherical, we deduce that the nanoparticle's melting temperature is inversely proportional to $N^{-\frac{1}{3}}$, where $N$ is the number of atoms in the nanoparticle:
    \begin{equation}
        T_{m,\text{NP}} = T_{m,\text{bulk}} - aN^{-\frac{1}{3}}
    \end{equation}
    where $a = T_{m,\text{bulk}} \times C\left(\frac{\pi}{6}\right)^{\frac{1}{3}}$.

    This relationship is valid for large nanoparticles. In fact,  experiments have shown that this relation does not hold for small nanoparticles \cite{mpnp-baletto, mpnp-calvo}. With a higher number of surface atoms compared to the bulk configuration, their behavior significantly deviates from that of the infinite system, making it unlikely that a straightforward extrapolation of macroscopic values will accurately predict their properties. Thus, this study exclusively considers nanoparticles with sizes greater than 1000 atoms.

    Above 693 K, bulk cobalt adopts a face-centered cubic structure. Therefore, to avoid shape effects and to be consistent with the determination of the melting temperature using the biphased cell method, the nanoparticles used in this study are perfect truncated octahedra containing 1289, 2406, 4033, 6266, and 9201 atoms, which correspond to the magic numbers of that shape. These nanoparticles are equilibrated at 900 K, then heated up to 2000 K for 13.75 ns. We chose to split the trajectory in intervals of 62.5 ps at constant temperature, every 5 K, to be sure that the energy is converged. Thus, the overall heating rate is 0.08 K.ps\textsuperscript{-1}. For these calculations, the system evolves through Langevin dynamics. The melting temperature corresponds to the maximum of the first derivative of the energy with respect to temperature, which was well defined in all cases, as shown in Fig. \ref{fig:mpnp_cv}.

    Extrapolation to infinitely large particles using the linear fit of the nanoparticles' melting points, as illustrated in Fig. \ref{fig:mp}, yields a bulk melting point of 1785 K, a mere 15 K higher than the experimental melting point. This outcome underscores the present q-SNAP's exceptional ability to accurately capture the dynamics of cobalt nanoparticles. It implicitly also demonstrates that the DFT-PBE level of theory captures the free energy difference between solid and liquid Co remarkably well.

    \section{\label{sec1:ccl}Conclusion}
    In this article we presented the development, validation, and application of a q-SNAP to predict static and dynamic properties of bulk Co, surfaces, and nanoparticles.
    
    The training set consisted of a total of 1049 cobalt structures, encompassing diverse crystallographic lattices with varying degrees of deformation, submitted to molecular dynamics at various temperatures and pressures, as well as surfaces and nanoparticles deliberately displaced from their equilibrium states to capture a wide range of atomic environments.
    
    While the computational cost associated with obtaining the training set may seem significant, it is quickly offset by the new capabilities gained once the q-SNAP is operational. This potential enables simulations of several dozens of nanoseconds, involving thousands of atoms, while closely maintaining the precision of DFT-PBE. The present q-SNAP reproduces remarkably well atomic vibrations (phonon dispersions) compared with results from DFT-PBE calculations, thus enabling accurate predictions of quantities such as thermal expansion, the hcp-fcc phase transition temperature, and the melting point. The values obtained for these properties with q-SNAP are in remarkable agreement with experimental data, thus implicitly validating the accuracy of the underlying DFT-PBE level of theory.
    
    Given its precise reproduction of DFT surface energies, one expects its aptitude in correctly modeling nanoparticles, which indeed is demonstrated in the present work. It accurately reproduces static properties such as the relative stability of nanoparticles of different shapes as a function of their size. Calculations of the vacancy formation energy on vertices of nanoparticles and related subtle structural changes demonstrate the high sensitivity of the present q-SNAP. Its robustness is further illustrated by accurately extrapolating melting temperatures of nanoparticles to the bulk value as a function of the number of atoms.
    
    This diverse range of capabilities positions the q-SNAP as a promising tool for catalysis applications, particularly due to its versatility in accommodating additional elements of various types—an advantage not shared by EAM potentials, which are restricted mostly to metals.
    
    The DFT calculations of the training set use a spin-polarized Hamiltonian. In all of the present configurations, the spins are ordered ferromagnetically. This facilitates the correlation between the structural environment and SNAP descriptors, which lack explicit information on magnetic moments. For other transition metals with lower Curie temperatures, such as iron, accounting for magnetic ordering becomes important. The development of machine-learned potentials including magnetism has been demonstrated for pure Fe \cite{drautz-iron}. The extension to systems with multiple magnetic elements is a field of ongoing and leading research. 
    
    It is important to note that the present q-SNAP was specifically constructed to model hcp and fcc cobalt, along with their surfaces and nanoparticles in various structures. Caution must be exercised when employing machine-learned potentials to simulate systems significantly different from the training set.
    
    In conclusion, the present work demonstrates that current machine-learned potentials are able to reach near DFT accuracy at the speed of conventional interatomic potentials, accurately capturing subtle dynamic effect such as the structure, energetics, and dynamics of surface atoms on nanoparticles. 
    
    The newly developed potential used in the present work can be downloaded from Ref. \cite{potential}.

    \begin{acknowledgments}
    We would like to thank Dr. Farka\v{s} from Cardiff University and Dr. de Leeuw from Utrecht University for providing the structural data of small cobalt nanoparticles \cite{farkas} used in our training set. Their contribution was greatly appreciated.

    Finally, we would like to thank all colleagues at Materials Design, especially Xavier Rozanska, for their help and support of this work.
    \end{acknowledgments}

    \bibliography{Co_paper}

\begin{thebibliography}{75}
\bibitem [{pot()}]{potential}%
  \BibitemOpen
  \href@noop {} {}\bibinfo {howpublished} {The {URL} will be active once the
  paper is published.}\BibitemShut {Stop}%
\end{thebibliography}
\end{document}